\documentclass[proof]{WileyASNA-v1}
\UseRawInputEncoding

\articletype{Article Type}%

\received{26 April 2016}
\revised{6 June 2016}
\accepted{6 June 2016}

\begin{document}

\title{Describing the evolution and perturbations to biodiversity using a simple dynamical model}

\author[1]{L. G. Bar\~ao}

\author[1]{L. M. de S\'a}

\author[1]{A. Bernardo}

\author[1]{J. E. Horvath}

\authormark{L. G. Bar\~ao \textsc{et al}}

\address[1]{\orgdiv{Universidade de S\~{a}o Paulo}, \orgname{Instituto de Astronomia, Geof\'{i}sica e Ci\^{e}ncias Atmosf\'{e}ricas}, \orgaddress{\city{S\~{a}o Paulo}, \state{SP}, \country{Brazil}}}

\corres{*L.G. Bar\~ao \email{lucasgbarao@usp.br }}
\presentaddress{R. do Mat\~{a}o, 1226 - Cidade Universit\'{a}ria, 05508-090, S\~{a}o Paulo-SP, Brazil}

\abstract{In this work, we outline a mathematical description of biodiversity evolution throughout the Phanerozoic based on a simple coupled system of two differential equations and on the division of genera in two classes - Short-lived and Long-lived types, as used by Rohde and Muller. We show that while the division in only two classes cannot capture the complexity of biodiversity evolution in short scales, broad trends can be described well. We also compute the systems' Green functions as a way to qualitatively describe the possible effects of intense perturbations of astrophyisical origin on the subsequent biodiversity evolution. }


\jnlcitation{\cname{%
\author{L. G. Bar\~ao}, 
\author{L. M. de S\'a}, 
\author{A. Bernardo}, and
\author{J. E. Horvath}} (\cyear{2023}), 
\ctitle{Describing the evolution and perturbations to biodiversity using a simple dynamical model}, \cvol{}.}


\maketitle

\section{Introduction}

``Hard sciences'' developed and expanded after the Scientific Revolution with an increasing interest in a {\it mathematical description} of reality, with a rich and long history \cite{Kline} tracing back to the Greek rationalism of Classical Antiquity, and its strong commitment to finding a formal formulation of physical problems \cite{Thomas}. On the other hand, work on ``Life science'' problems have tended to hold a different viewpoint, and a general mathematization of life phenomena is, in contrast to physical phenomena, not present. The Newtonian model of the type \textit{general mathematical law }$ \rightarrow$\textit{ application to specific case} does not hold the same paradigmatic status, and has even been challenged \cite{Turchin}. It is true that there is a body of mathematical tools and models in Life sciences, but they are mostly derived from phenomenology and do not play the same role as the broad fundamental laws in Hard sciences. Thermodynamics and a few other examples are exceptions to this rule, and over the years an interface of mathematical Biology/Ecology has been closing the gap between both approaches.

Attempts to understand the behavior of biological systems from a mathematical framework, ultimately formulating a set of ``laws'' inductively have been pursued, for example, in the study of population dynamics. The work of \citet{Lotka}, \citet{Volterra} and many others \cite{others} are well-known and have been elaborated for a long time. A general Lotka-Volterra type equation for population dynamics relates the rate of change of the number of individuals of one population to the population itself, and other ambient/external features that contribute to it. This kind of formalism is simple and flexible, and a variety of known biological features that determine the evolution of a given population can be included in it. 

It was only quite recently that this established framework was criticized and reformulated \cite{GC} to move away from its Aristotelian view of physical dynamics (in Lotka-Volterra the rate of change is a {\it first order} derivative and, therefore, is only sensitive to instantaneous variations). What followed was the formulation of a {\it second order} formalism to deal with the (second) temporal variation of the rate of change, introducing naturally the idea of inertia, within a ''Galilean turn'', i.e., as in the history of physics, there is a break with old first-order concepts and a search for second-order models \cite{GC, Clark, Guinzburg, Yee}. Being a richer structure, a second order formalism passed tests of recorded evidence of population dynamics that were difficult to fit within Lotka-Volterra schemes, and offers a new perspective of several observed phenomena.

The problem we wish to solve is not quite related to populations, but to the issue of biodiversity. It is usual to identify the latter with the number of genera $N_{G}$ living at a given time (the attempt to identify a deeper level of species is very difficult and leads to considerable error when the identification in the fossil record is performed). 

Inspired by the second-order approach, we have investigated a description of genera over geological time of the latter inertial type. We believe that the assumptions are reasonable, and the validation of such a model would be achieved by a comparison with the actual record, while the insights that will rise from the analysis of the dynamical system can be inductively formulated and tested. A first attempt to engulf this problem within a single, oscillator-type differential equation was inconclusive \cite{Review}, but allowed the development presented and discussed in this work. We note that this kind of work has an important input from astronomical events: nearby supernovae \cite{Fields} and gamma-ray bursts \cite{GH} are the type of cosmic catastrophes that would perturb the biosphere and prompt large changes, sometimes extinctions \cite{Brian}. In any case, all kinds of sudden perturbations, in the sense of being very short compared to the typical timescales of the biosphere evolution $1-10\text{ Myr}$ can be modeled as non-homogeneous abrupt functions in a dynamical system evolving with time. This is the idea of a Green function, which contains the response of the system to a sudden perturbation. As is stands, there is a large potential gain in the mathematical formulation of the biodiversity evolution.

This work is structured as follows: Section \ref{sec:single_eq} reviews the fundamental single equation for the evolution of $N_G$ constructed and analyzed before \cite{Horvath2014}. Section \ref{sec:double_spring} will motivate and present the new framework, a kind of ``zeroth order'' one to deal with the global biodiversity of the biosphere in terms of Short-lived and Long-lived genera, to be characterized in detail. Section \ref{sec:green} deals with the solutions (including the Green function) and their relevance to the existing fossil record. The conclusions and perspectives of this work will be presented in Section \ref{sec:conclusions}.

\section{Single-equation second order model}
\label{sec:single_eq}

With the idea of an inertial, Galilean description of the evolution of biodiversity, a second-order differential equation was motivated, constructed and solved in \citet{Horvath2014}. This simple approach had as a dynamical variable the total number of biological genera $N_{G}$, and at a linear level reads

\begin{equation}
\label{eq:ng}
\frac{d^{2}N_{G}}{dt^{2}} + \beta\frac{dN_{G}}{dt} + \alpha{\bigl( {N_{G}-N_{G}^{max}} \bigr)} = f(N_{G}, {\dot{N_{G}}}, t) .
\end{equation}

The second and third terms represent different effects on $N_{G}$. The second corresponds to a modeling of the ecological ``friction'' (i.e. how the growth behaves when the biodiversity was growing or decreasing at a certain rate), which introduces a characteristic damping time $\tau = 1/\beta$, especially relevant to the recovery after a large perturbation. The third is analogous to the {\it carrying capacity} in Lotka-Volterra systems, in which $\alpha \leq 0$ for consistency. Finally, the r.h.s. contains the perturbing ``forces'', and will be zero or very small in steady states, but include Dirac deltas at the time of large, sudden perturbations.

A finer analysis of the physical/biological nature of each of these terms is still lacking, although the general form of Equation \ref{eq:ng} is general enough for a linear description. However, it is not guaranteed that a single linear differential equation can describe all the features observed in the biodiversity curve \cite{RS}. In fact, in \citet{Horvath2014} the stationary solutions and Green function were derived, and while the inertial behavior is trivially accommodated, the aftermath of the well-studied Permo-Triassic extinction \cite{P-T} could not be well reproduced by the Green function behavior. This fact led \cite{Review} to declare the model as ''inconclusive''. It is apparent that the complexity of the recovery after a large perturbation is not captured in Eq. \ref{eq:ng}. In addition, there is no obvious {\it internal} mechanism that can induce a sudden flip of the solutions behavior, as many biologists/ecologists/paleontologists believe. For them, the biosphere does not in general suffer external (astrophysical) perturbations, but rather its internal dynamics should be behind the observed extinctions \cite{Bond}. The K-T event may be an exception showing that cosmic external factors could be involved \cite{Alvarez}. It is apparent that the model must be revisited and reformulated to attempt a proper mathematical description of the $N_{G} (t)$ curve. This is the task we undertake in the present work.


\section{A Double Spring Model for the temporal behavior of genera}
\label{sec:double_spring}

A widely used proxy for the temporal behavior of the evolution of biodiversity in the biosphere is the compilation by Sepkoski \cite{RS}, comprising the Phanerozoic Marine diversity. This important work has, however, some important problems that have been discussed over the years. These have mainly concerned possible biases in the data - what are real effects and what are observational artifacts. This motivated a considerable amount of work in analyzing the data and developing methods to correct it (like \cite{Alroy},\cite{Bush} and others). Still, even after such corrections, the overall shape of the Sepkoski curve tends to be preserved, along with its large scale trends. We thus employ the original Sepkoski data to develop and test the methods presented below, which, as we will show, are best employed as descriptions of those large scale trends. In any case, we describe each step in general terms, such that they could be applied to any other diversity curve, likely with similar results.

\begin{figure}[ht!]
    \centering
    \includegraphics[width=0.9\columnwidth]{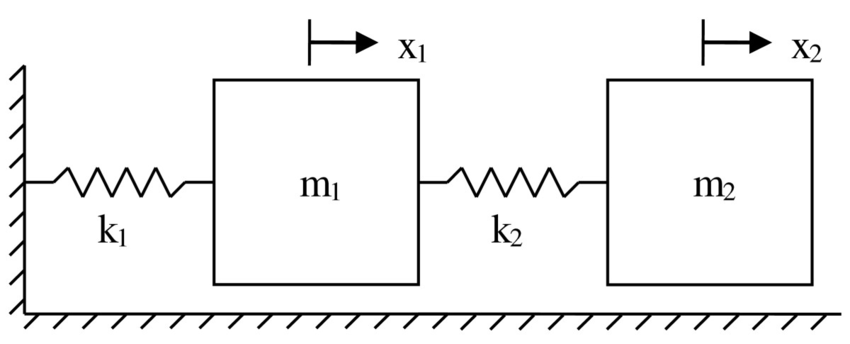}
    \caption{A Double Spring system representation \citep{Dumond}.}
    \label{fig:mass-string}
\end{figure}

As a proof-of-concept system for the modeling of the temporal evolution of biodiversity, identified with the number of genera $N_{G}$ as a function of time, we set a system of coupled differential equations for two masses $m_1$ and $m_2$ connected to each other by a spring of constant $k_2$, while a spring of constant $k_1$ connects $m_1$ to a wall (Figure \ref{fig:mass-string}). Friction and air resistance are not included, so that the dynamical equations are

\begin{align}
    \label{eq:x1}
    \frac{d^2 x_1}{d t^2}(t) &= -\frac{k_1}{m_1}(x_1(t) - r_1) + \frac{k_2}{m_1}(x_2(t) - x_1(t) - r_2), \\
    \label{eq:x2}
    \frac{d^2 x_2}{d t^2}(t) &= -\frac{k_2}{m_2}(x_2(t) - r_2),
\end{align}

\noindent where $x_i$ is the position of mass $m_i$, and $r_i$ its equilibrium position. We then identify the mass displacements $x_i$ with the number of genera ${N_{Gi}}(t)$ over time belonging to two different classes ${N_{G1}}$ and ${N_{G2}}$, representing Short-lived genera and Long-lived genera, with respect to the survival time of the genera. This separation has been suggested several times in the literature \citep{RM,Mellot2011,Gilinsky}, and could explain the behavior observed in the data. For consistency, we impose that

\begin{equation}
    \label{eq:sc_ng_sum}
    N_{G1}(t) + N_{G2}(t) = N_{G}(t),
\end{equation}

\noindent where $N_{G}(t)$ is a function describing the Sepkoski curve (Section \ref{sec:solving}). Although in principle it is possible to try to model {\it any} two-class division of genera within the Sepkoski data, we shall always refer to the Long-lived and Short-lived genera classes defined above. The main reason is that they can be readily associated with the physical parameters of the double spring-mass system.

We can now proceed to characterize the biological analogue of the masses $m_i$ as {\it genera inertia}, i.e., a measure of how sensitive each genera class is to changes by external pressures. The larger the inertia of a genera class, the more slowly its diversity ($N_{Gi}$) responds to changing environmental conditions or interaction with other genera. In the discussion of this issue, \citet{GC} present two possible effects that could be interpreted as the inertia of populations: the Maternal Effect and the Predator-Prey mechanism (in this latter case, appropriate in terms of populations, not genera). Whatever the ultimate cause of inertia is, and based on the work of \cite{RM}, the Short-lived genera could be defined by a {\it smaller} inertia compared with the Long-lived genera.

Continuing the discussion of the model, we state that the spring constants $k_i$, in turn, represent the biological forces that drive the genera towards an equilibrium state, which we label as "internal forces". These are ``internal'' to the biological system, including populations and environment, in contrast to ``external forces'' originating from {\it outside} of the environment (i.e. a meteorite impact or any other astrophysical perturbation). In this case, a $N_G$ below equilibrium means that there is a relative abundance of resources and/or less competition for the available ecological niches. Therefore, the chances of a genus to survive is greater, and there are more opportunities for genera creation, driving biodiversity upwards. On the other hand, a number of genera above equilibrium means a relative scarcity of resources, and more competition for individual niches, leading to the extinction of some genera and a downward trend in biodiversity. 

The interplay between resource availability and competition naturally results in the existence of an equilibrium state under fixed environmental conditions, such that the rate of genera creation is equal to the rate of extinction, and the overall number of genera in each class stays the same. This means that the equilibrium positions $r_i$ have the most clear and immediate meaning, as the equilibrium point for each $N_{Gi}$.

A different aspect of the interpretation of the parameters is to consider their dimensions. While $N_{Gi}$ and $r_i$ can be naturally taken as the dimensionless counts of the number of genera within a class, the $k_i$, which in the mechanical system has dimensions of $mass \times time^{-2}$, depends on how we define the measurement of biological inertia, which is not totally clear. We can, however, define the parameters $\tau_{ij}$ as 

\begin{equation}
    \label{eq:def_tau}
    \tau_{ij} = \sqrt{\frac{m_i}{k_j}},
\end{equation}

\noindent which have dimensions of time. We can then rewrite Eqs. \ref{eq:x1} and \ref{eq:x2} as 

\begin{align}
    \label{eq:x1_tau}
    \frac{d^2{N_{G1}}}{{dt^2}}(t) &= -\tau_{11}^{-2}(N_{G1}(t) - r_1) + \tau_{21}^{-2}(N_{G2}(t) - N_{G1}(t) - r_2), \\
    \label{eq:x2_tau}
    \frac{d^2 N_{G2}}{dt^2}(t) &= -\tau_{22}^{-2}(N_{G2}(t) - r_2),
\end{align}

\noindent and work only with the characteristic times $\tau_{11}$ and $\tau_{22}$ of the populations $N_{G1}$ and $N_{G2}$, respectively, which provide a ready way to identify the Long-lived and the Short-lived genera, and can also be compared to the general idea of a genus ``eigenperiod'', a concept defined by \cite{GC} as the natural period of oscillation of the genera, when subjected to a ``biological force''.

\subsection{Solving the system of equations}
\label{sec:solving}

\begin{figure}
    \centering
    \includegraphics[width=\columnwidth]{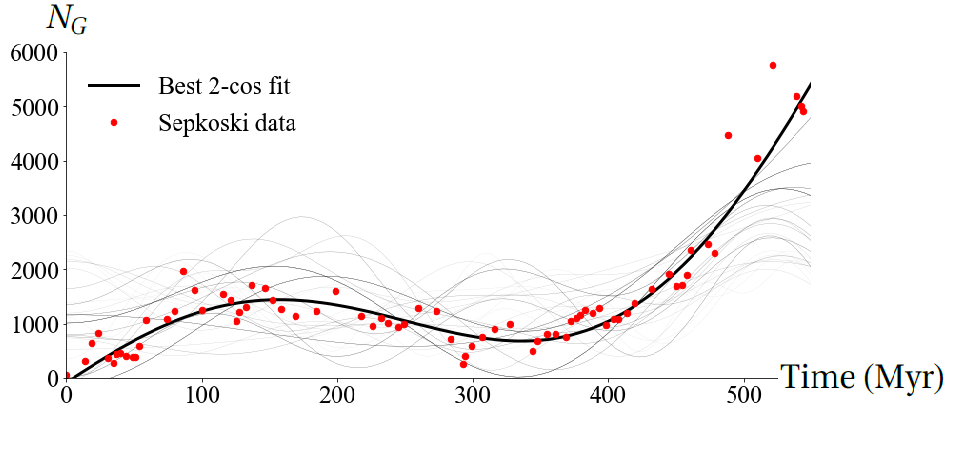}
    \caption{The Sepkoski data (red points) overlaid on the results of our fitting procedure. The light gray lines represent the ensemble of fits generated by our $R^2$ with annealing method. The best fit, and the one we employ here, is shown by the black line, with parameters indicated in Table \ref{tab:sc_cos_fit}.}
    \label{fig:sepkoski}
\end{figure}

The most general solution of any dynamical system can be written as the linear combination of its normal modes of oscillation, which in the case of the $n$ spring-mass system are $n$ cosine functions with different (natural, or resonant) frequencies. For our $n=2$ system, this means that we can at most write $N_{G1}(t)$, $N_{G2}(t)$ and, consequently, $N_{G}(t)$, as  

\begin{align}
    \label{eq:ng1_coss}
    N_{G1}(t) &= \alpha_0 + \alpha_1\cos{\left(\omega_1(t + \phi)\right)} + \alpha_2\cos{\left(\omega_2(t + \phi)\right)},\\
    \label{eq:ng2_coss}
     N_{G2}(t) &= \beta_0 + \beta_1\cos{\left(\omega_1(t + \phi)\right)} + \beta_2\cos{\left(\omega_2(t + \phi)\right)},\\
     \label{eq:ng_coss}
     N_{G}(t) &= \mathrm{a}_0 + \mathrm{a}_1\cos{\left(\omega_1(t + \phi)\right)} + \mathrm{a}_2\cos{\left(\omega_2(t + \phi)\right)},
\end{align}

\noindent with parameters determined by a $R^2$ goodness-of-fit test to the Sepkoski data. Because we find the fit to be sensitive to the initial guess, we employ an annealing method by iterating the fit a number of times, with each iteration taking the results from the previous one, plus a random noise, as a starting guess. We also add a time shift $\phi$ as an additional time parameter to allow for a phase shift. This guarantees that we find a global best-fit, which is shown in Figure \ref{fig:sepkoski}; the resulting parameters are shown in Table \ref{tab:sc_cos_fit}.

\begin{center}
\begin{tabular}{lr}
\toprule
Parameter & Value \\
\midrule
$a_0$ ($N_G\cdot\mathrm{Myr}$) & 4676.30 \\
$a_1$ ($N_G\cdot\mathrm{Myr}$) & 3125.83 \\
$a_2$ ($N_G\cdot\mathrm{Myr}$) & 6674.45 \\
$\omega_1$ ($\mathrm{Myr}^{-1}$) & 0.009 \\
$\omega_2$ ($\mathrm{Myr}^{-1}$) &  0.004 \\
$\phi$ ($\mathrm{Myr}$) & 267.567 \\
\bottomrule 
\end{tabular}
\captionof{table}{Best parameters for the Sepkoski curve fit.}
\label{tab:sc_cos_fit}
\end{center}

The resulting $N_{G}(t)$ fit thus defines the two natural frequencies our system {\it must} have to provide the best possible match to the Sepkoski data. Solving its eigenvalue equation for the $\tau_{ij}$ and defining $\tau:=\tau_{11}$ yields

\begin{align}
    \label{eq:tau_bound}
    \frac{1}{\omega _1} 	&< \tau < \frac{1}{\omega _2},\\
    \label{eq:tau22}
    \tau_{22} &= \frac{1}{\tau \omega _1 \omega _2},\\
    \label{eq:tau12}
    \tau_{12} &= \tau \sqrt{\frac{1}{\left(\tau^2 \omega _1^2-1\right) \left(1-\tau^2 \omega _2^2\right)}}.
   \end{align}

Once the $\tau_{ij}$ is well determined, we can then solve Equations \ref{eq:ng1_coss} and \ref{eq:ng2_coss} and impose Equation \ref{eq:sc_ng_sum}. This procedure fully determines the $N_{Gi}$ amplitudes as

\begin{align}
    \label{eq:alpha_defs}
    \alpha_0 &= \frac{1}{2} \left(a_1-r_2\right), & \alpha_1 &= \frac{a_2 \left(\tau ^2 \omega _2^2-1\right)}{2 \tau ^2 \omega _2^2-1}, & \alpha_2 &= \frac{a_2 \omega _1^2 \left(\tau^2 \omega _2^2-1\right)}{\omega _2^2 \left(2 \tau^2 \omega _1^2-1\right)}, \\
    \label{eq:beta_defs}
    \beta_0 &= \frac{1}{2} \left(a_1+r_2\right), & \beta_1 &= \frac{a_2 \tau ^2 \omega _2^2}{2 \tau ^2 \omega _2^2-1}, & \beta_2 &= \frac{a_2 \tau ^2 \omega _1^2}{2 \tau ^2 \omega _1^2-1}.
\end{align}

\begin{figure}
    \centering
    \includegraphics[width=\columnwidth]{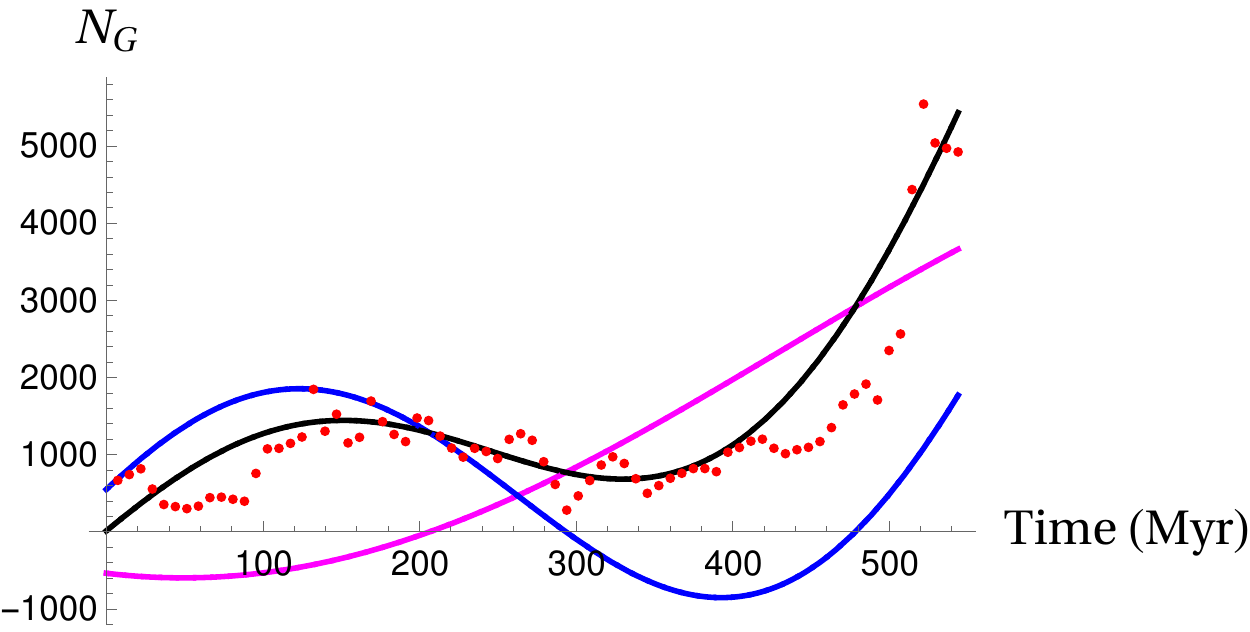}
    \caption{A possibility of division between Short-lived and Long-lived genera. The magenta, blue and black lines represent, respectively, the Long-lived genera, Short-lived genera and the best fit to the Sepkoski curve $N_{G}$. The limitation of this model implies in ``negative'' numbers of genera in some points. Even so, as presented in \cite{RM}, the Short-lived oscillate more over time than Long-lived ones.}
    \label{fig:ShortLong}
\end{figure}

Equations \ref{eq:tau_bound}-\ref{eq:beta_defs} fully define $N_{G1}(t)$ and $N_{G2}(t)$, with their sum always equal to $N_{G}(t)$, while keeping $\tau$ as a free parameter within the limits of Equation \ref{eq:tau_bound}. As $\tau$ and $\tau_{22}$ are the characteristic times of populations $N_{G1}$ and $N_{G2}$, respectively, this allows for some freedom in ``distributing'' the natural frequencies $\omega_1,\omega_2$ between the two genera classes. The extreme cases, where one population has characteristic time $1/\omega_1$ and the other $1/\omega_2$, nicely points towards a Long-lived and Short-lived genera distinction, although limitations of the model provide an unphysical negative number of genera in some intervals (Figure \ref{fig:ShortLong}) and the values of the $\omega_i$ are in excess of expectations. We discuss this feature in the next section. 

\subsection{Fourier Transform approach}
\label{sec:fourier}

\begin{figure}
    \centering
    \includegraphics[width=\columnwidth]{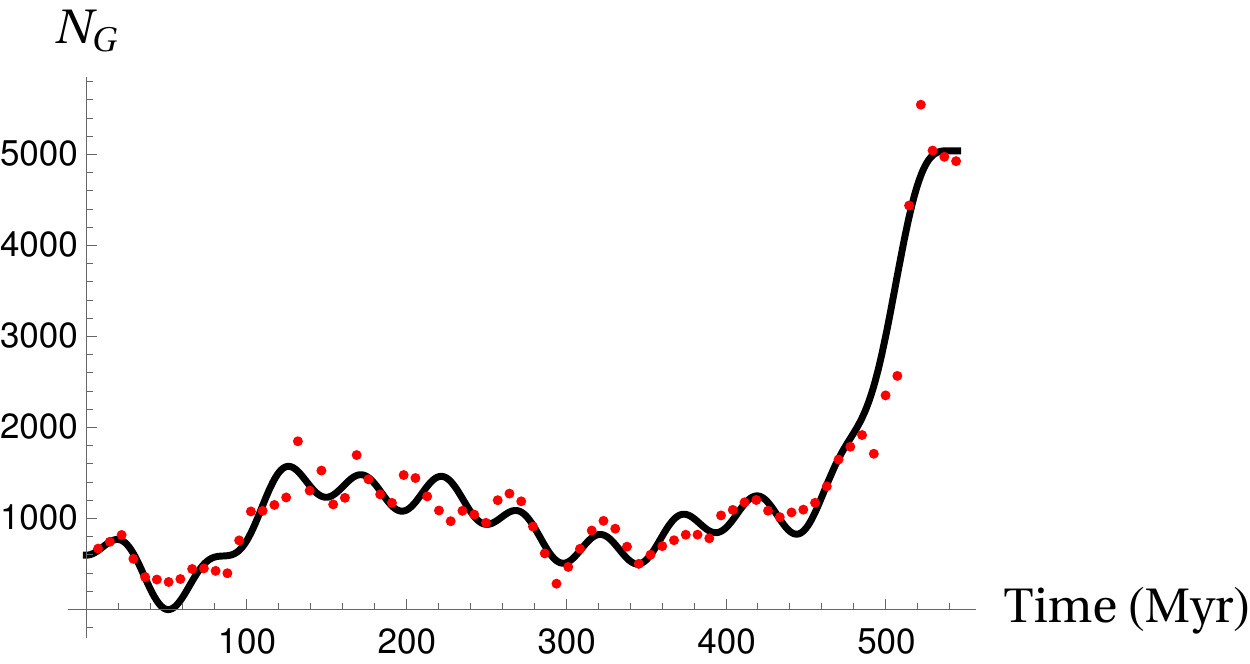}
    \caption{The Sepkoski curve (red dots) with a fit based on Fast Fourier Transform and using the eleven most relevant amplitudes.}
    \label{fig:Fourier}
\end{figure}

A more efficient method to describe the biodiversity Curve $N_{G}(t)$ is to use a Fourier Transform. As our system of equations only supports a sum of two cosines as a solution, we cannot employ a Fourier series as a solution, but we briefly investigate it as a comparison to our method. For this purpose, we apply a Fast Fourier Transform to the Sepkoski data, and extract the eleven most important frequencies (highest amplitudes in the transform). Table \ref{tab:FFT} shows these amplitudes and their respective periods, which are generally one order of magnitude smaller than those obtained for our fit (Table \ref{tab:sc_cos_fit}). These differences rise because the two most important frequencies from the Fourier transform do \textit{not} necessarily provide the best fit of the simpler two-cosine model; instead, the best fit captures mainly large-scale behavior. For a system with enough degrees of freedom, on the other hand, both results should quickly converge as the general solution becomes a longer sum of cosines. Figure \ref{fig:Fourier} overlays on the data the cosine sum with the amplitudes and frequencies from \ref{tab:FFT}.

The two periods obtained from the two-cosine fit for the Sepkoski curve (\ref{tab:sc_cos_fit}) - $T_1=112.5\text{ Myr}$ and $T_2=242.7\text{ Myr}$ are an order of magnitude greater from the patterns found in other works, such as $T = 62\text{ Myr}$ \cite{Mellot2011} and $T = 26\text{ Myr}$ \cite{RS2}. But, as stated above, our simple system has a low resolution, and thus it is hard to identify any periodicity hidden in the curve besides the most general trends. However, as seen in Table \ref{tab:FFT} above, these two periods (approximately) appear with high amplitudes ($57\text{ Myr}$ and $29\text{ Myr}$), as expected from those works.

\begin{center}
\begin{tabular}{cc}	
  \toprule
            \textbf{Module of amplitudes}	& \textbf{Periods}\\
            {$N_G\cdot\text{Myr}$} & {Myr}\\
            \midrule
$1282$          & constant         \\
$901$           & $57$  \\
$770$         & $173$  \\
$528$           & $87$  \\
$401$           & $29$  \\
$343$           & $43$  \\
$340$           & $35$  \\
$317$          & $22$  \\
$210$          & $25$  \\
$186$          & $8$  \\  
$131$           & $17$  \\
        \bottomrule
	\end{tabular}
	\captionof{table}{The eleven highest amplitudes and their respective frequencies, computed with FFT and applied to the Sepkoski curve.\label{tab:FFT}}
\end{center}

To increase this resolution while keeping an ordinary linear differential equations model, it is thus necessary to consider a larger number of genera classes. Given that the Sepkoski curve has 75 data points, the FFT results in $75$ frequencies in the best scenario, where none of them are discarded as noise. Then the system with best possible accuracy must have 75 different genera types. However, this is a mathematical statement only, since the physical division \cite{RM} of genera into Short-lived and Long-lived classes is motivated by the raw data, while a large number of classes is not. Another possibility is to set up a non-linear system of differential equations that have more normal modes of oscillation, but keep the number of genera classes. 

\section{Perturbations on the system and mass extinctions}
\label{sec:green}

The study of the response to a sudden (in a geological timescale) external force of any dynamical system that seeks to describe biological populations is motivated by events such as activity in Large Igneous Provinces (LIPs) with global effects \cite{Bond}; and in astronomical terms, a massive minor body collision, supernovae explosions or $\gamma$-ray bursts, among other events. Traditional physical mathematical techniques offer a ready method to calculate this response in the form of a Green function, which is the solution of the system for a Dirac delta external force, and which we find for the system presented here.

\begin{figure}[ht!]
    \centering
    \includegraphics[width=\columnwidth]{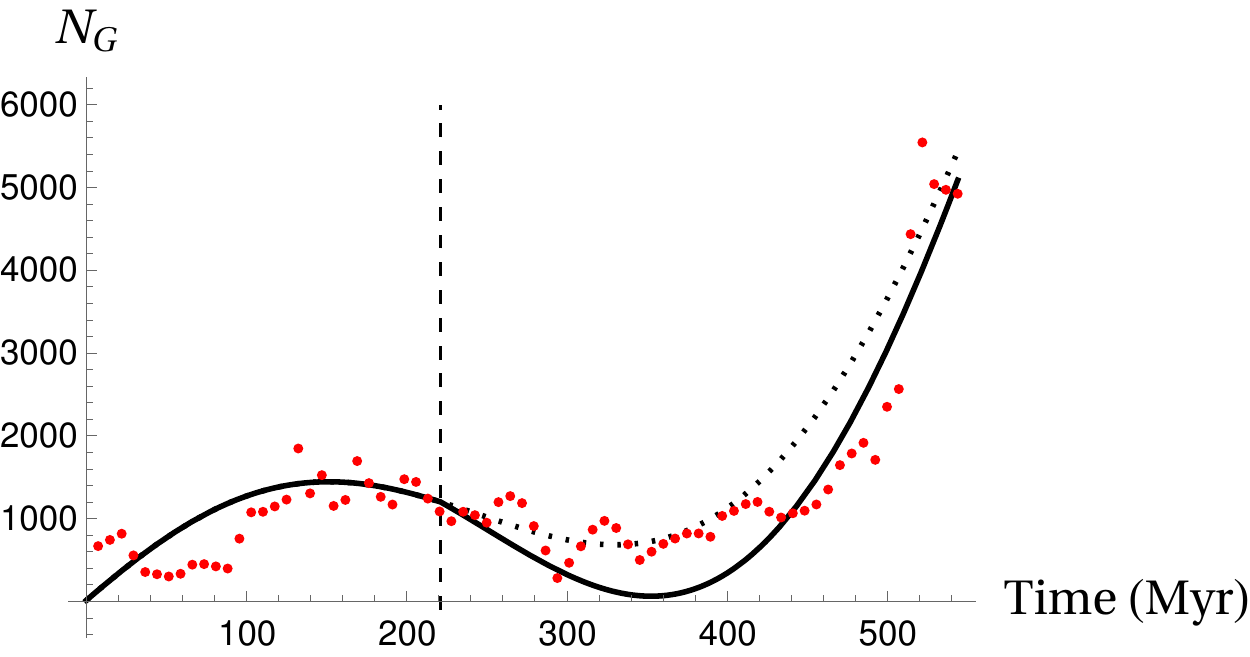}
    \caption{Simulations of external forces acting on the Sepkoski curve. When the perturbation occurs at a time of natural decline, the extinction is intensified. The vertical dashed and dotted lines shows, respectively, the moment of perturbation and the normal Sepkoski curve. The black line shows the Sepkoski curve after the "kick".}
    \label{fig:kicks1}
    \centering
    \includegraphics[width=\columnwidth]{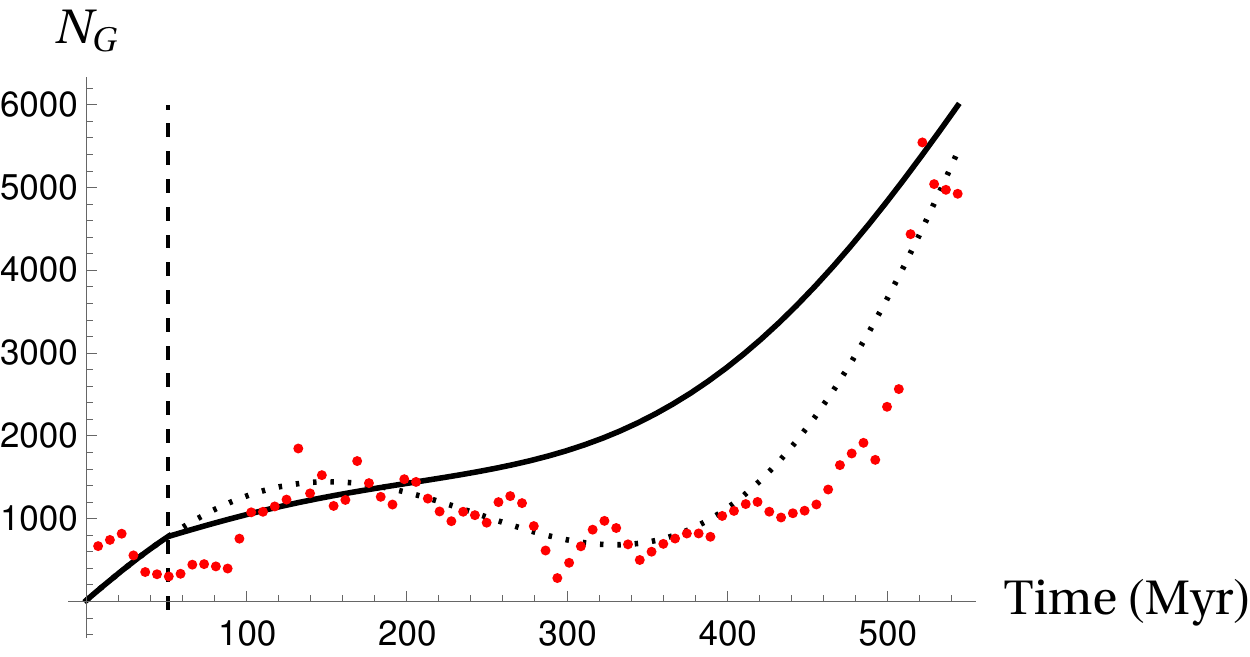}
    \caption{Simulations of external forces acting on the Sepkoski curve. When the perturbation occurs at a time of natural growth, the genera are slightly decreased only. The vertical dashed and dotted lines shows, respectively, the moment of perturbation and the normal Sepkoski curve. The black line shows the Sepkoski curve after the "kick".}
    \label{fig:kicks2}
\end{figure}

We add a Delta $\delta (t -t_{0})$ to the r.h.s of Equations \ref{eq:x1_tau} and \ref{eq:x2_tau}, and find the associated Green function to the Sepkoski fit providing the boundary conditions at $t<t_0$. Thus, considering that the perturbation in the system happens at a certain time $t_{0}$, the already known $N_{G}(t)$ holds before the perturbation, while after it the $N_{G}(t)$ function will adopt a behavior that is closely related to the characteristics of the two genera types, and their tendencies at the time of the ``kick''. Figures \ref{fig:kicks1} and \ref{fig:kicks2} show two different possibilities of ``kicks'', at different times and their effect on the the biodiversity curve. 

Analyzing the behavior of the number of genera when subjected to a sudden intense external force (the ``kick''), we can infer a plausible and expected tendency: when the disturbance happens at a time when the number of genera are naturally decreasing, the decline is intensified and the number of genera tends to zero. However, when the disturbance happens in the opposite situation, the number of genera are slightly decreased only. In the example shown in Figure \ref{fig:kicks1}, the curve nearly reaches $N_G=0$, and can cross the axis for an intense enough perturbation. Although the system cannot capture the fine behavior of genera evolution, it illustrates that the response of biodiversity to catastrophic events can be highly sensitive to trends already present at the moment of such events.

This solution thus points toward a trend on why and when genera can become extinct. For the genera that are more sensitive to external forces (low inertia) or that are caught in a decreasing moment being already affected by a hostile environment, the sudden perturbation can possibly drive these genera to extinction. On the other hand, if the genera are more resilient (high inertia), the probabilities of being extinct are much lower. This qualitative result fits well with recent studies that have indicated that an increased biodiversity makes populations more resilient to extinction events \citep{Weeks}. There is also a fortuitous element playing an important role in this sketchy extinction picture that should be further explored. In the present model the recovery is very long, due to its simplicity, while \citet{KW}, after a careful analysis of the biodiversity curve which took into account effects such as possible biases in the fossil record, determined the existence of a characteristic "recovery time" of $\sim 10\text{ Myr}$ which marks the delay between extinction and the following recovery events, and which was shown to hold even for background extinction events. As discussed in Section \ref{sec:fourier}, such short scales could still be modeled within the general steps of our method, but only for a more complex system, which, for example, allows for more degrees of freedom, or admits non-linear interactions between populations.

\section{Conclusions}
\label{sec:conclusions}

We seek in this work to provide a formalism to describe the evolution of biodiversity through the Phanerozoic. Using a coupled dynamical model, that allows us to explore the hypothesis of Short-lived and Long-lived genera, we have obtained relevant solutions and calculated the Green function, which is useful to simulate events of mass extinction caused by external perturbations. As seen in the previous sections, using a simple system of differential equations as a model is sufficient to describe the long-term behavior of genera over time, having a division of genera in two types in mind (the Short-lived and Long-lived ones). We characterized their differences, with the introduction of the parameter $\tau$ (analogous to the ``eigenperiod'' as defined by \cite{GC}) for each genera. With the latter, the definition of these genera classes suggested, for example, in \cite{RM} are supported, and it is seen that the Short-lived genera oscillate over geological time, while the Long-lived ones evolve but do not even oscillate.  

Furthermore, the application of the Green function to the system proved to be a useful tool to study the response of the biodiversity to the ``external forces'', including possible mass extinctions mechanisms. Even the simplest coupled differential equation model can thus reveal insights about what can influence the survival of genera, such as their own properties, like the sensitivity to environmental chances, but also the state of biodiversity at the moment of the perturbation. This means, for example, that if a nearby supernovae or a meteorite impact hit the planet during a period of high diversification, its consequences could be dampened. However, if the disturbance happens at a time of decline, then a mass extinction is highly likely. Nevertheless, this simple model is quantitatively unable to explain the recovery times found by \cite{KW}, although the qualitative recovery features should hold.

It is clearly that use of a simple system of differential equations (Equations \ref{eq:x1} and \ref{eq:x2}) is not sufficient to describe the details seen in experimental curves of number of genera over time (the Sepkoski curve), the latter a sum of the two genera classes. Thus, two options are left. To continue preserving the two genera division it is necessary to resort to systems of non-linear differential equations or even partial differential equations. Alternatively, if the linearity of the solution is to be preserved, then it becomes necessary to consider a division of genera in more than two classes. Whatever the option, we must think of this problem within a mathematical description of biological phenomena that should be more complex.  


\section*{Acknowledgments}

Financial support was provided by the \fundingAgency{Funda\c{c}\~ao de Amparo \`a Pesquisa do Estado de S\~{a}o Paulo (FAPESP)}, grants \fundingNumber{13/26258-4} and \fundingNumber{2020/08518-2}. The \fundingAgency{Conselho Nacional de Desenvolvimento Cient\'ifico e Tecnol\'ogico (CNPq)} is acknowledged by JEH for the award of a Research Fellowship; and by LMS acknowledges for financial support, grant \fundingNumber{140794/2021-2}.


\begin{thebibliography}{999}

\bibitem[Alroy et al. (2001)]{Alroy}
Alroy, J. et al. (2001), {\it PNAS}, {\it 98}, 6261

\bibitem[Alvarez (1987)]{Alvarez}
Alvarez, L.W., Alvarez, W., Asaro, F. and Michel, H.V. (1980), {\it Science}, {\it 208}, 1095

\bibitem[Bond et al. (2014)]{Bond}
Bond, D. P., Wignall, P. B., Keller, G., and Kerr, A. C. Large igneous provinces and mass extinctions: an update, In: Volcanism, impacts, and mass extinctions: causes and effects {\it 505}, 29, 2014

\bibitem[Bond and Grasby (2017)]{Bond}
Bond, D.P.G. and Grasby, S.E. (2017), {\it Palaeogeography, Palaeoclimatology, Palaeoecology},{\it 478}, 3

\bibitem[Burgess, Bowring and Shen (2001)]{P-T}
Burgess, S.D, Bowring, S. and Shen, S.-Z. (2014), {\it PNAS}, {\it 111}, 3316 

\bibitem[Bush et al. (2004)]{Bush}
Bush, A.M., Markey, M.J. and Marshall, C.R. (2004), {\it Paleobiology}, {\it 30}, 666

\bibitem[Clark (1971)]{Clark}
Clark, G.P. (1971), {\it Ecology}, {\it 52}, 606

\bibitem[Domund and Baddour (2014)]{Dumond}
Dumond, P. and Baddour, N. (2014). {\it SpringerPlus}, {\it 3}, 1

\bibitem[Fields et al. (2020)]{Fields}
 Fields, B. D., Melott, A. L., Ellis, J., Ertel, A.F., Fry, B.J., Lieberman, B.S., Liu, Z., Miller, J.A.  and Thomas, B.C. (2020), {\it PNAS}, {\it 17}, 21008

\bibitem[Galante and Horvath (2007)]{GH}
Galante, D. and Horvath, J.E. (2007), {\it Inter. Jour. Astrobiology}, {\it 6}, 19

\bibitem[Gilinsky (1988)]{Gilinsky}
Gilinsky, N. L. (1988),{\it Paleobiology}, {\it 14}, 370

\bibitem[Guinzburg (1986)]{Guinzburg}
Ginzburg, L.R.  (1986), {\it J. Theor. Biol.}, {\it 122}, 385

\bibitem[Guinzburg and Colyvan (2004)]{GC}
Guinzburg, L.R. and Colyvan, M. {Ecological Orbits: How Planets Move and Populations Grow}, Oxford University Press: Oxford, UK, 2004

\bibitem[Horvath (2014)]{Horvath2014}
Horvath, J.E. (2014), {\it Challenges}, {\it 5}, 324

\bibitem[Kingsland (1985)]{others}
Kingsland, S.E. { Modeling Nature: Episodes in the History of Population Ecology}, 1st ed., University of Chicago Press: Chicago, IL, USA, 1985.

\bibitem[Kirchner and Weil (2000)]{KW}
Kirchner, J.W. and Weil A. (2000), {\it Nature}, {\it 404}, 177

\bibitem[Kline (1981)]{Kline}
Kline, M. {Mathematics and the Physical World}, Dover Publications: UK, 1981

\bibitem[Lotka (1925)]{Lotka}
Lotka, A.J. Elements of Physical Biology , Williams \& Wilkins Company: Baltimore, MD, USA, 1925

\bibitem[Mellot et al. (2011)]{Mellot2011}
Melott, A. L. and Bambach, R. K. (2011), {\it Paleobiology}, {\it 37}, 383

\bibitem[Raup and Sepkoski (1982)]{RS}
Raup, D. M., and Sepkoski Jr, J.J.(1982), {\it Science}, {\it 215}, 1501

\bibitem[Raup and Sepkoski (1984)]{RS2}
Raup, D.M. and Sepkoski Jr, J.J. (1984), {\it PNAS}, {\it 81}, 801

\bibitem[Rohde and Muller (2005)]{RM}
Rohde, R. A. and Muller, R. A. (2005), {\it Nature}, {\it 434}, 208

\bibitem[Sudakow et al. (2022)]{Review}
Sudakow, I., Myers, C., Petrovskii, S., Sumrall, C.D. and Witts, J. (2022), {\it Physics of Life Reviews}, {\it 41}, 1

\bibitem[Thomas (1939)]{Thomas}
Thomas, I. {Greek Mathematical Works: Thales to Euclid},Volume I ; Harvard University Press: Cambridge, MA, USA, 1939.

\bibitem[Thomas et al. (2004)]{Brian}
Thomas, B., Melott, A., Lieberman, B., Laird, C., Martin, L., Medvedev, M., Cannizzo, J., Gehrels, N. and  Jackman, C. Did a gamma-ray burst initiate the late Ordovician mass extinction? American Physical Society, April Meeting, 2004, May 1-4, 2004, Denver, Colorado April 2004, MEETING ID: APR04, abstract id. S10.014

\bibitem[Turchin(2001)]{Turchin}
Turchin, P. (2001), {\it Oikos}, {\it 94}, 17

\bibitem[Volterra (1926)]{Volterra}
Volterra, V. (1926), {\it Nature},{\it 118}, 558

\bibitem[Weeks et al. (2022)]{Weeks}
 Weeks, B.C., Naeem, S., Lasky, J.R. and Tobias, J.A. (2022), {\it Ecology Letters}, {\it 25}, 697

 \bibitem[Yee (1980)]{Yee}
Yee, J. A. (1980), {\it Theor. Pop. Biol.}, {\it 18}, 175


\end{thebibliography}

\section*{Author Biography}
(if applicable)

\begin{biography}{\includegraphics[width=60pt,height=70pt,draft]{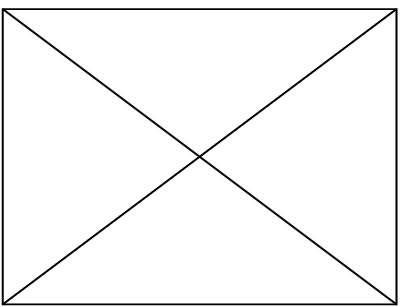}}{\textbf{J.E. Horvath} graduate (1985) and Ph.D (1989) in Physical Sciences at the Universidad Nacional de La Plata (La Plata, Argentina). Tenured Professor at the IAG-USP  since 1992 and Full Professor since 2022. Founder and Director of the {\it N\'ucleo de Pesquisas em Astrobiologia} at USP 2011-2021. His interests span Relativistic Astrophysics, High Energies, Cosmology, Astrobiology, Education and Philosophy of Science, contributing regularly to them for more than 30 years. Author of 8 didactic books for Graduate, Undergraduate and general public on these subjects. Class I Researcher of the CNPq Federal Council, Brazil. .}
\end{biography}

\end{document}